\documentclass[conference]{IEEEtran}
\IEEEoverridecommandlockouts
\usepackage{cite}
\usepackage{amsmath,amssymb,amsfonts}
\usepackage{algorithmic}
\usepackage{graphicx}
\usepackage{setspace}
\usepackage{textcomp}
\usepackage{caption}
\usepackage{xcolor}
\def\BibTeX{{\rm B\kern-.05em{\sc i\kern-.025em b}\kern-.08em
    T\kern-.1667em\lower.7ex\hbox{E}\kern-.125emX}}

\begin{document}

\title{EVAN: Evolutional Video Streaming Adaptation via Neural Representation\\
}
\author{\IEEEauthorblockN{Mufan Liu\IEEEauthorrefmark{1}, Le Yang\IEEEauthorrefmark{2}, Yiling Xu\IEEEauthorrefmark{1}, Ye-kui Wang\IEEEauthorrefmark{3}, Jenq-Neng Hwang\IEEEauthorrefmark{4}}
\IEEEauthorblockA{\IEEEauthorrefmark{1}\textit{CMIC, Shanghai Jiao Tong University, China} \\
\IEEEauthorrefmark{2}\textit{Department of Electrical and Computer Engineering, University of Canterbury, New Zealand } \\
\IEEEauthorrefmark{3}\textit{Bytedance Inc., USA; }
\IEEEauthorrefmark{4}\textit{Department of Electrical and Computer Engineering, University of Washington, USA}\\
\{\{sudo\_evan, yl.xu\}@sjtu.edu.cn, le.yang@canterbury.ac.nz, yekui.wang@bytedance.com, hwang@uw.edu\}\thanks{This paper is supported in part by National Natural Science Foundation of China (62371290, U20A20185) and 111 project (BP0719010). The corresponding author is Yiling Xu(e-mail: yl.xu@sjtu.edu.cn).}}}

\maketitle
\begin{abstract}
Adaptive bitrate (ABR) using conventional codecs cannot further modify the bitrate once a decision has been made, exhibiting limited adaptation capability. This may result in either overly conservative or overly aggressive bitrate selection, which could cause either inefficient utilization of the network bandwidth or frequent re-buffering, respectively. Neural representation for video (NeRV), which embeds the video content into neural network weights, allows video reconstruction with incomplete models. Specifically, the recovery of one frame can be achieved without relying on the decoding of adjacent frames. NeRV has the potential to provide high video reconstruction quality and, more importantly, pave the way for developing more flexible ABR strategies for video transmission. In this work, a new framework, named Evolutional Video streaming Adaptation via Neural representation (EVAN), which can adaptively transmit NeRV models based on soft actor-critic (SAC) reinforcement learning, is proposed. EVAN is trained with a more exploitative strategy and utilizes progressive playback to avoid re-buffering. Experiments showed that EVAN can outperform existing ABRs with 50\% reduction in re-buffering and achieve nearly 20\% improvement in users' quality of experience (QoE).
\end{abstract}
\begin{IEEEkeywords}
Neural Representation, Adaptive Bitrate, Video on Demand, Video Streaming, Model Pruning
\end{IEEEkeywords}

\section{Introduction}
Streamed videos nowadays dominate mobile data consumption. Meanwhile, there has been a steady increase in demands for higher video quality. According to \cite{cisco}, by 2023, 66 percent of the newly installed flat-panel TVs will be ultra-high definition (UHD), compared with a value of 33 percent in 2018. However, Internet video transmission often experiences considerable fluctuations and unpredictability in network conditions, which renders guaranteeing the delivered video quality a significant challenge.

Adaptive bitrate (ABR) techniques were developed to enhance the quality of the received video under time-varying network conditions. These methods are executed on client-side video players. They dynamically choose a bitrate for each video chunk. ABR makes bitrate decisions on the basis of various measurements such as the estimated network throughput and playback buffer occupancy. The objective is to maximize users' quality of experience (QoE) by adapting the video bitrate according to different network conditions. Most ABR algorithms can be broadly divided into two categories, namely rule-based and learning-based approaches. Rule-based methods use pre-ﬁxed rules for selecting bitrates based on buffering and/or thoughput observations \cite{mpc, festive}. They require signiﬁcant tuning and do not tradeoff between different QoE metrics. Learning-based ABR like Pensieve \cite{Pensieve}, on the other hand, applies reinforcement learning (RL) to learn a control policy for bitrate adaptation. 
\begin{figure*}
    \centering
    \includegraphics[width=0.7\textwidth]{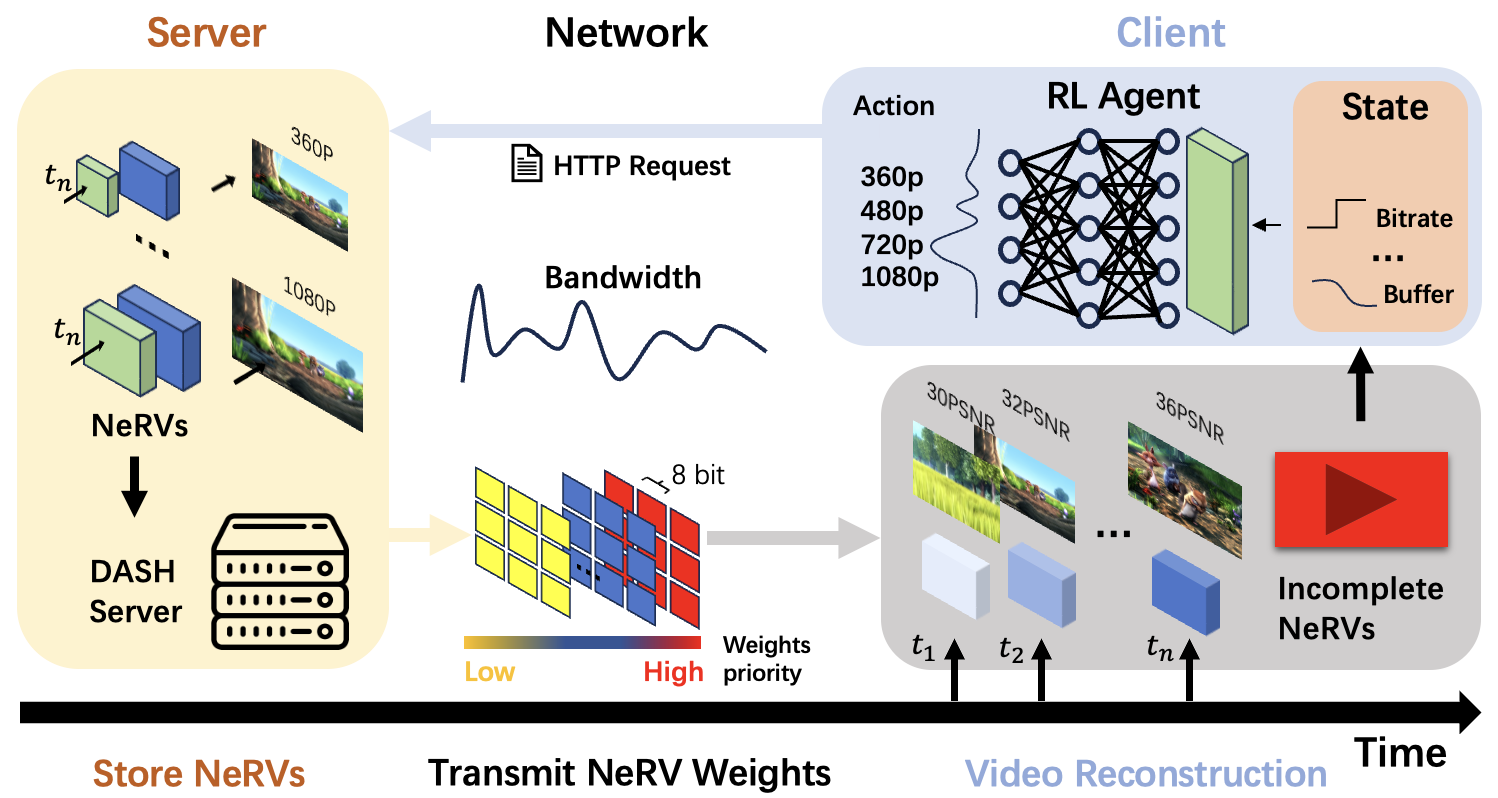}
    \caption{Framework of EVAN. The server stores NeRV weights for different representations of video chunks. NeRV weights are transmitted according to their priority. The client can simultaneously reconstruct the video while downloading the associated NeRV weights. The RL agent decides the next chunk's representation to be transmitted, based on the observed network state}
    \label{framework}
 \end{figure*}
The aforementioned ABR algorithms, nevertheless, all employ a monolithic encoding strategy, where typically each video chunk is encoded separately for each quality level. Moreover, the decision on fetching a segment of a certain quality is considered final, once it is made by an ABR algorithm based on the estimated network capacity. To achieve finer-grained bitrate control, recent work developed improved ABR through resorting to the emerging progress in video coding and processing. Specifically, 
\cite{Yeo} proposed the transmission of lower-quality video representations, which are then quality-enhanced using a client-side super-resolution (SR) model. Following a similar approach, \cite{SR} added colorization before carrying out SR, which leads to further decrease in video size thanks to transmitting grey-scale videos only. \cite{swift} separated video into distinct layers for transmission via layered coding. These approaches increase, to some extent, the granularity of video bitrate decisions while conserving network bandwidth, but the granularity of video enhancement is still restricted by the performance of the SR model and layered coding. 

Neural Representations for Video (NeRV) \cite{NeRV} has been shown to be an effective generative video encoding and decoding framework. It is capable of producing high-quality video reconstructions with continuous model pruning. 
Its `continuous' pruning ratio enables more precise quality enhancement in a progressive manner than most existing scalable video coding approaches \cite{SHEVC}. Recognizing that NeRV has not yet been integrated into current video streaming systems, we propose a new technique, named Evoluational Video streaming Adaptation via Neural representation (EVAN), to achieve more flexible bitrate decisions with minimum re-buffering. The motivations behind this work are as follows. First, NeRV can provide higher video quality over traditional codecs within the video on demand (VoD) bits per pixel (BPP) range. Additionally, NeRV allows for simultaneous video playback during model downloading, thus mitigating re-buffering issues under unfavorable network conditions. Finally, its independent frame recovery ability allows users to begin playback from \textit{any} frame, prompting a smoother transition. Our contributions in this work include:
\begin{itemize}
    \item A new framework, EVAN, for attaining adaptive video streaming using NeRV is presented. It is able to perform adaptive selection of a NeRV model to download for video reconstruction under varying nework conditions.
    \item We include a progressive playback strategy that enables downloading the NeRV model while simultaneously reconstructing the video from an incomplete NeRV model.
    \item We systematically evaluate EVAN against state-of-the-art ABR methods in terms of the re-buffering gain and user QoE improvement.
\end{itemize}
\section{EVAN}
The proposed EVAN framework utilizes NeRV for content delivery on the existing Dynamic Adaptive Streaming over HTTP (DASH) \cite{dash}. EVAN involves training NeRV models with the representations of each video chunk and storing them on the DASH server (see Fig. \ref{framework} for an illustration). The client evaluates the current network condition and decides on the appropriate bitrate for the next chunk, and then sends an HTTP request to the server. The DASH server transmits the requested chunk’s NeRV weights for the client to achieve video reconstruction and playback. The client is able to reconstruct a lower-quality version of the video based on the incomplete model already received, thus avoiding re-buffering under poor network conditions. In the following subsections, we detail the modules of EVAN.
\subsection{NeRV}
NeRV refers to neural representations for videos based on generative models. Instead of treating videos as a series of frames, NeRV models each segment of a video using a unique neural network. NeRV's encoding process involves training a neural network so that it overfits the given video, while the decoding process entails feeding the sequence number of frames into the neural network for forward processing. As a result, we can represent a video $V$ as $V=\{v_k\}_{k=1}^{T}$, where $v_k = f_\theta(k)$ implies frame $k$ being represented as a network $f$ parameterized by $\theta$, and $T$ is the number of frames in $V$. 
\subsection{Progressive Playback}
In DASH, the video cannot be played until the entire chunk is downloaded. If the next video chunk has not been downloaded before it is supposed to be played, re-buffering will occur. We found that most re-buffering occurs at the beginning of the video (see Fig. \ref{duration}), since each time a new video is started, the buffer is cleared, making the earlier parts of the video more prone to re-buffering. Meanwhile, viewers’ interests are heavily influenced by their experience at the beginning of the video consumption. Hence, if there is excessive initial re-buffering, the viewers are likely to skip the whole video. In such cases, prioritizing an early start for video playback, even at the cost of sacrificing a certain level of initial quality, may be more meaningful than applying a longer initial re-buffering to achieve higher initial quality. Therefore, it is important to have progressive playback during the downloading process to speed up the video start and reduce re-buffering time.
\begin{figure}[!b]
 \centering
 \includegraphics[width=0.8\linewidth]{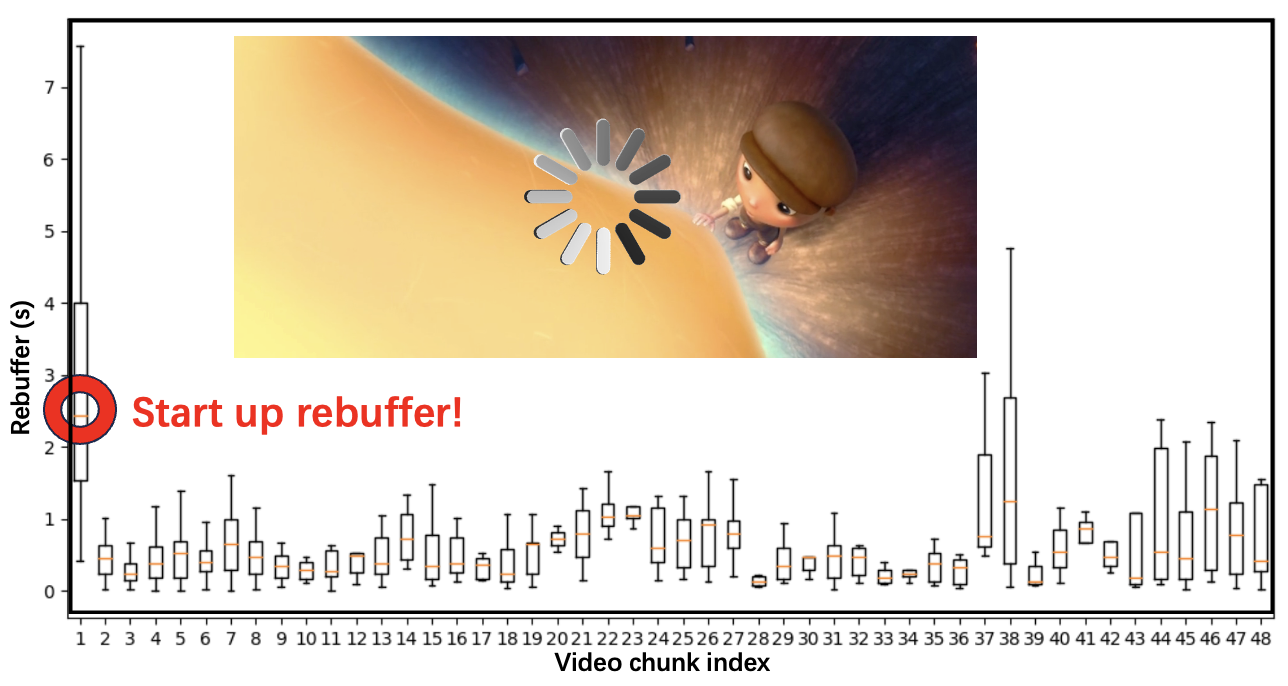}
 \caption{Re-buffering duration over chunks}
 \label{duration}
\end{figure}
The decoding procedure of NeRV is essentially the same as the forward processing of a trained neural network, allowing us to reconstruct the video in real time as we receive the network weights. In order to realize the progressive playback strategy mentioned above, it is necessary to prioritize the weights according to their significance before sending them. The goal would be to maximize the quality of the video that will be played first before the model is fully downloaded. We employ the global non-structural pruning method to determine the importance of network weights. Specifically, weights are organized according to their cumulative probability density (CDF), meaning that weights with higher probability mass will be assigned higher priority. Moreover, all the weights are quantized using 8 bits, thus avoiding extra overhead for notifying startup for different resolutions. \cite{NeRV} demonstrated that compressing the model weights from 32 bits to 8 bits has a negligible effect on the quality of the reconstructed video. This brings significant reduction in the model size without evident degradation of video quality. For practical ABR implementations, it is essential to set a minimum threshold on the downloaded model size for video playback. Videos will only be played when the downloaded model size surpasses this threshold. We will discuss the selection of this threshold in the performance evaluation section (Section III).
\subsection{Quality Assessment Model}
To fairly compare the performance of EVAN against existing ABR techniques, we compute the QoE of chunk $i$ as follows as in \cite{mpc}
\begin{equation}
    \label{QoE}
    \text{QoE}_i = q\left(R_i\right)-\mu T_i- \left|q\left(R_{i+1}\right)-q\left(R_i\right)\right|.
\end{equation}
where $q\left(R_i\right)$ is the `equivalent' encoding bitrate of the NeRV model for chunk $i$, and $T_i$ is the re-buffering time needed to download the NeRV for chunk $i$. The last term in (\ref{QoE}) penalizes variations in quality between adjacent video chunks. 

In our implementation, for each representation of a video chunk, we trained a NeRV model of the \textit{same} size. To quantify subjective video quality, we used $q\left(R_i\right)$ to map a received NeRV model to the encoding bitrate in terms of a conventional codec H. 264 with bitrate $R_i$. We set the BPP for NeRV and H. 264 to be equal to $\beta_i$ and define $q\left(R_i\right)$ as
\begin{equation}
    q\left(R_i\right) = R_{\rm V} = R_{i}\frac{log(f_{\rm V}(\beta_i)/\gamma_0)}{log(f_{0}(\beta_i)/\gamma_0)}.
    \label{ratemap}
\end{equation}
where $f_{\rm V}(\cdot)$ denotes the BPP-peak signal to noise ratio (PSNR) mapping for NeRV, while $f_0(\cdot)$ finds the PSNR from BPP for H. 264, both of which are established using numerical fitting (see e.g., \cite{waterloo}). $\gamma_0$ is a tuning parameter. In other words, the equivalent encoding bitrate $q\left(R_i\right)$ is proportional to the model size $R_i$ and the NeRV coding `gain'. From (\ref{ratemap}), we can see that the coding gain depends roughly on the ratio of the logarithms of the PSNRs from NeRV and H.264. The use of the logarithms takes into consideration the fact that the improvement in the human visual quality becomes less evident when the PSNR further increases. 


Video quality degradation could occur when playing a video based on an incompletely downloaded NeRV model. This will be reflected by modifying the bitrate mapping $q\left(R_i\right)$ to account for model sparsity. Specifically, we record the PSNRs of the pruned models trained in \cite{waterloo} and relate them to the model sparsity levels. It was found that their relationship can be accurately modeled using an exponential function, which is given by
\begin{equation}
\label{sparsity}
    \gamma_{\beta_i}(r_t) = f_{\rm V}(\beta_i) - be^{c(1 - r_t)}.
\end{equation}
Here, $r_t\in [0,1]$ represents the model sparsity level at time $t$, and $b,c$ are fitting parameters. 

For the considered adaptive video transmission problem, $r_t$ becomes equal to $r_t=\frac{\int_{t_0}^t B_{\tau}d\tau}{d(R_i)}$, where $B_{t}$ is the network bandwidth at time $t$, $t_0$ is the starting time point for downloading video chunk $i$, and $d(R_i)$ is the size of chunk $i$. Thus, substituting the definition of $r_t$, and combining (\ref{ratemap}) and (\ref{sparsity}), we can write the \textit{instantaneous} equivalent NeRV bitrate at time $t$, which is $R_{\rm V}(t)$, as
\begin{equation}
\label{instantenous}
    R_{\rm V}(t) = R_{i}\frac{log(\gamma_{\beta_i}((\int_{\tau = t_0}^{t} B_\tau d\tau)/d(R_i)))/\gamma_0)}{log(f_0(\beta_i)/\gamma_0)}.
\end{equation}
Note that the QoE for chunk $i$ becomes meaningful only after its playback starts at time $t_p$. Besides, the QoE would remain constant until the end of the playback, once the downloading of chunk $i$ is finished. Applying (\ref{instantenous}), the average equivalent encoding bitrate for chunk $i$ is thus equal to 
\begin{equation}
    R_{\rm V}' = \frac{1}{T} \int_{t_p}^{t_d} R_{\rm V}(\tau) d\tau + \frac{(T-(t_d-t_p))}{T} R_{\rm V}.  
\end{equation}
The first term comes from the process of downloading chunk $i$, which ends at time $t_d$, while the second term is from the playback spanning from $t_d$ to $t_p + T$, where $T$ is the chunk duration. 

The re-buffering time $T_i$ in (\ref{QoE}) is now computed as follows. If at the starting point of the playback $t_p$, the downloaded model size, which is $\int_{t_0}^{t_p} B_\tau d\tau$, is larger than a pre-specified threshold $r_0\cdot d(R_i)$ (i.e., $r_{t_p}\geq r_0$), no initial re-buffering is needed, as the playback could be initiated immediately. In this case, $T_i = 0$. Otherwise, initial re-buffering is required to fill the buffer until the download model size reaches the threshold. In this case, $T_i$ satisfies $\int_{t_0}^{t_p+T_i} B_\tau d\tau = r_0\cdot d(R_i)$.

\subsection{Soft Actor-Critic (SAC) Reinforcement Learning}
A variety of RL algorithms have been applied to solve complex ABR problems. In EVAN, we choose to use the soft actor- critic (SAC) approach \cite{SAC} since it has been shown to better trade-off between exploration and exploitation in training an agent. As a comparison, the A3C policy, due to its limited exploration capability, sometimes tends to make conservative bitrate selections in order to completely eliminate re-buffering \cite{SAC-PEN}. This approach does not benefit from NeRV’s progressive playback strategy as our goal is to choose the highest bitrate possible in EVAN with progressive playback.
\begin{figure}
 \centering
 \includegraphics[width=0.70\linewidth]{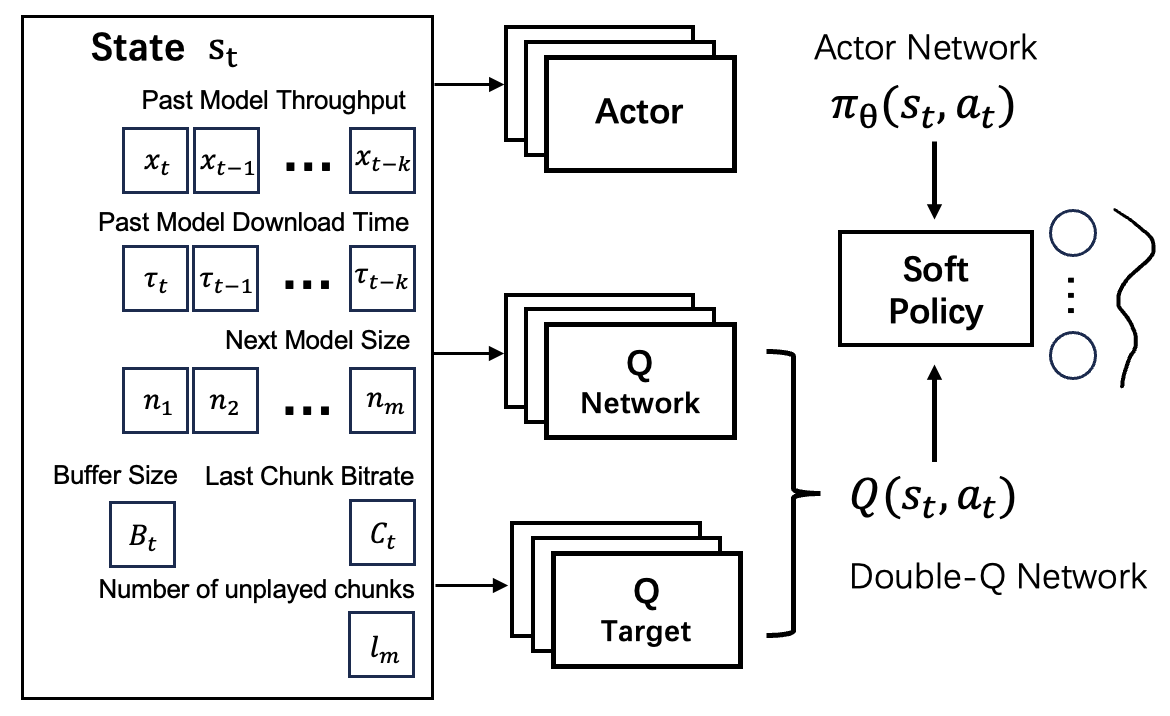}
 \caption{Structure of the soft actor-critic network}
 \label{fig:1}
 \end{figure}
Fig. \ref{fig:1} shows the SAC architecture, which includes an actor network for decision-making and two Q-networks for state-action value (Q) estimation. Weights between the two Q-networks are periodically alternated to prevent overestimation of the Q-values. Each Q-network consists of one evaluation network (Q-eval) and a target network (Q-target) to estimate the current and target Q values. SAC considers jointly maximizing the cumulative rewards and the entropy in the training goal. For each policy iteration, agents choose the optimal policy that maximizes the weighted sum of expected cumulative reward and entropy as follows
\begin{equation}
    \pi^* = \underset{\pi}{\rm \text{argmax}}
    \; \mathbb{E}_{a_t \sim \pi} r(s_t,a_t) + \beta \mathcal{H}(\pi(a_t|s_t)).
\end{equation}
We will not delve into the detailed derivation of the loss functions for the actor and Q networks in this paper. Readers can refer to \cite{SAC} for more details. The functionalities of SAC networks are explained below.
\begin{itemize}
\item Inputs: At the client side, after a NeRV model is downloaded at time $t$, the agent takes in the state input $s_t = (\boldsymbol{x_t},\boldsymbol{\tau_t},\boldsymbol{n_t},B_t, C_t, l_m)$, where $\boldsymbol{x_t}$ is the vector stores the past model downloading throughput, $\boldsymbol{\tau_t}$ stores the past model downloading time, $\boldsymbol{n_t}$ stores the available model sizes for the next chunk, $B_t$ is current buffer size, $C_t$ is last chunk's bitrate and $l_m$ is the number of remaining video chunks.
\item Actor network: The aim of the actor (policy) network is to minimize the Kullback-Leibler divergence (KLD) of the policy distribution and “Q shaped” probability mass distribution
\begin{small}
\begin{equation}
     J_\pi(\phi) =D_{\mathrm{KL}}(\pi_\phi\left(a_t \mid s_t\right) \| \exp (\frac{1}{\beta} Q_\theta\left(s_t, a_t\right)-\log Z\left(s_t\right))).
\end{equation}
\end{small}
where $J_\pi(\phi)$ is the actor network loss, $\pi_\phi$ is the actor policy, $Q_{\theta}$ represents the Q-eval network. $\beta$ is the weight of entropy and $\log Z(s_t)$ is the log partition function.
\item Q network: 
The Q network is used to approximate the state-action value, which is trained to maximize the advantage function $A(a_t, s_t)$ defined as
\begin{equation}
    A(a_t, s_t) = Q_{\theta}\left(s_t, a_t\right)-(r\left(s_t, a_t\right)+\gamma V\left(s_{t+1}\right)).
\end{equation}
As we have mentioned, Q-eval $Q_{\theta}$ is trained to estimate the current Q value $Q\left(s_t, a_t\right)$, while Q-target is trained to estimate the next state value using 
\begin{equation}
V\left(s_{t+1}\right) = Q_{\bar{\theta}}(s_{t+1}, a_{t+1})
-\beta \log (\pi_\phi\left(a_{t+1} \mid s_{t+1}\right).
\end{equation}
Thus, the training loss for the Q network is
\begin{small}
\begin{equation}
J_Q(\theta) = \mathbb{E}_{\pi_{\phi}}\left[\frac{1}{2}A(a_t, s_t)^2\right].
\end{equation}
\end{small}
\end{itemize}
\begin{figure}[!h]
 \centering
 \includegraphics[width=1\linewidth]{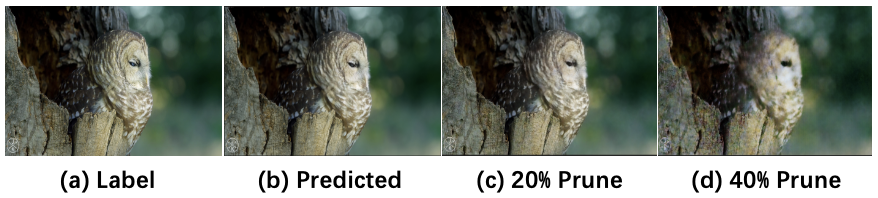}
 \caption{Reconstructed frames on different prune ratios}
 \label{duration}
\end{figure}
\begin{table*}
\centering
\caption{Tested video chunk size and NeRV models}
\resizebox{0.8\textwidth}{!}{
\begin{tabular}{c|c|c|c|c|c|c|c|c|c}
\hline\hline
Resolution & Bitrate & BPP & Avg. chunk size & NeRV size & Hidden Dim. & Feature map size & Lowest width & Channels & Upscale \\
\hline
360p & 750kbps &  0.137 & 0.40MB & 0.39MB & 128 & (8,12) & 48 & 10 & [5 3 3] \\

480p & 1750kbps & 0.174 & 0.90MB & 0.89MB & 128 & (8,12) & 72 & 10 & [5 3 2 2]\\

720p & 2350kbps & 0.086 &  1.20MB & 1.19MB & 256 & (9,16) & 70 & 15 & [5 2 2 2 2] \\

720p & 3000kbps & 0.107 & 1.50MB & 1.49MB & 256 & (9,16) & 86 & 15 & [5 2 2 2 2] \\

1080p & 4300kbps & 0.071 & 2.2MB & 2.19MB & 512 & (9,16) & 64 & 20 &[5 3 2 2 2]\\

1080p & 7000kbps & 0.112 & 3.5MB & 3.49MB & 512 & (9,16) & 94 & 28 & [5 3 2 2 2] \\
\hline
\end{tabular}}
\label{nerv model}
\end{table*}

\begin{figure*}[htbp]
\begin{minipage}{0.33\linewidth}
     \centering
     \includegraphics[width=0.9\linewidth]{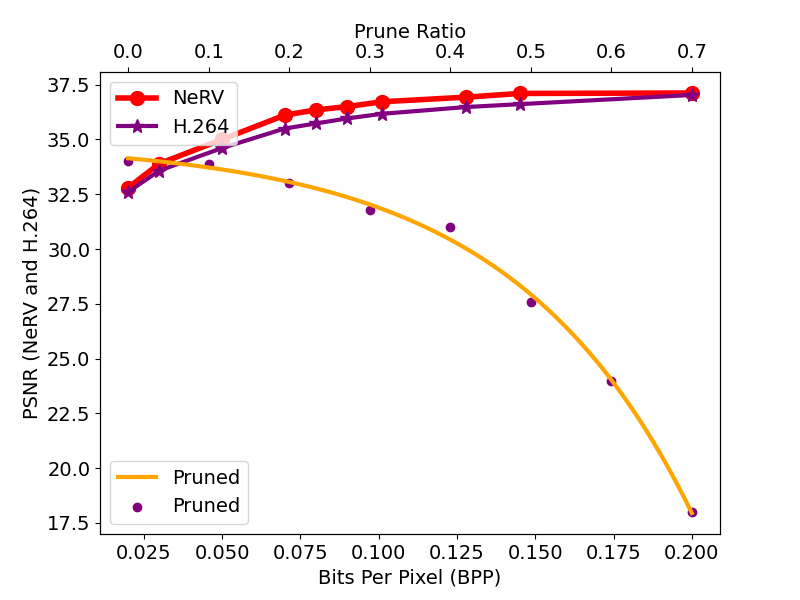}
     \caption*{(a)}
     \end{minipage}
     \begin{minipage}{0.33\linewidth}
     \centering
     \includegraphics[width=0.9\linewidth]{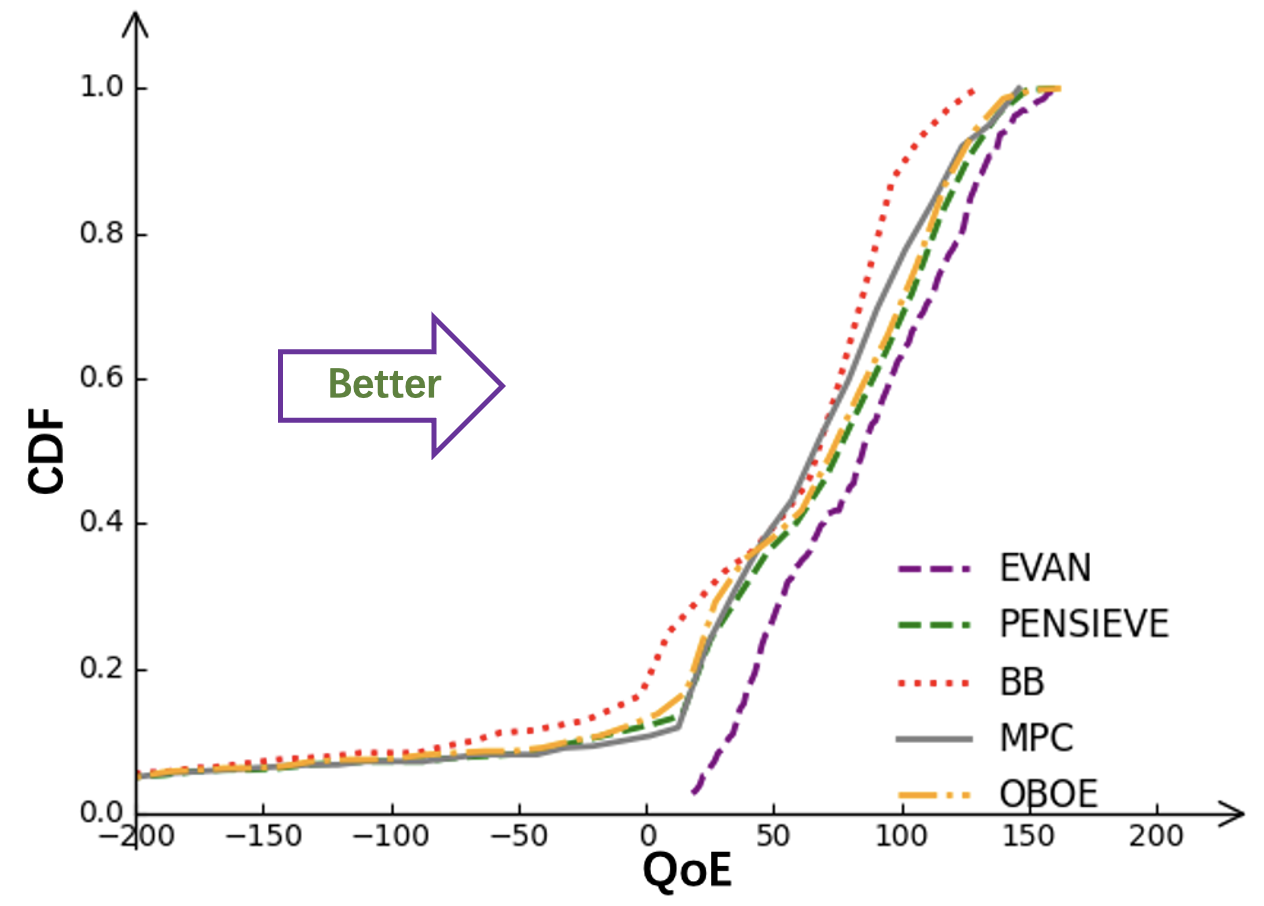}
     \caption*{(b)}
     \end{minipage}
     \begin{minipage}{0.33\linewidth}
     \centering
     \includegraphics[width=0.9\linewidth]{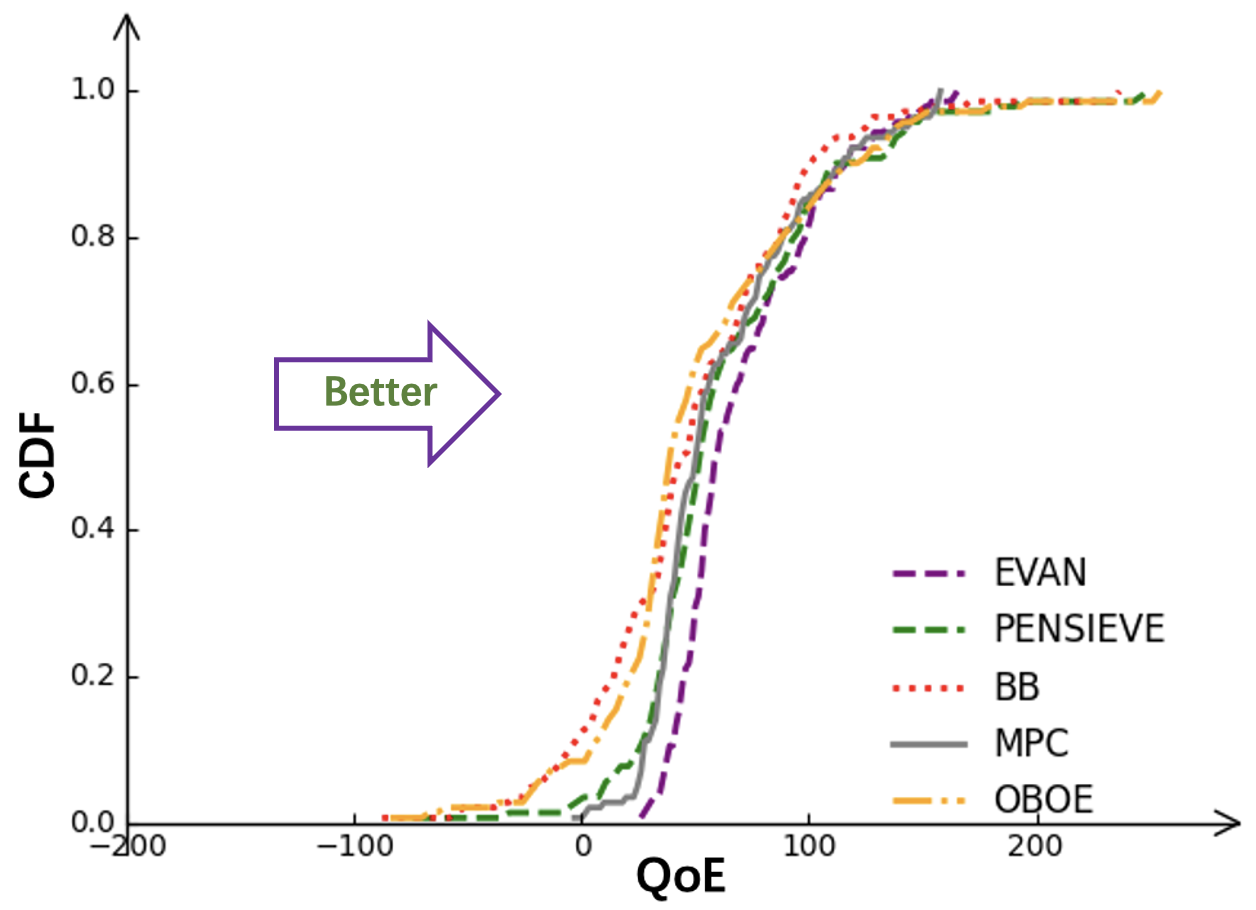}
     \caption*{(c)}
     \end{minipage}
     \begin{minipage}{0.5\linewidth}
     \centering
     \includegraphics[width=0.85\linewidth]{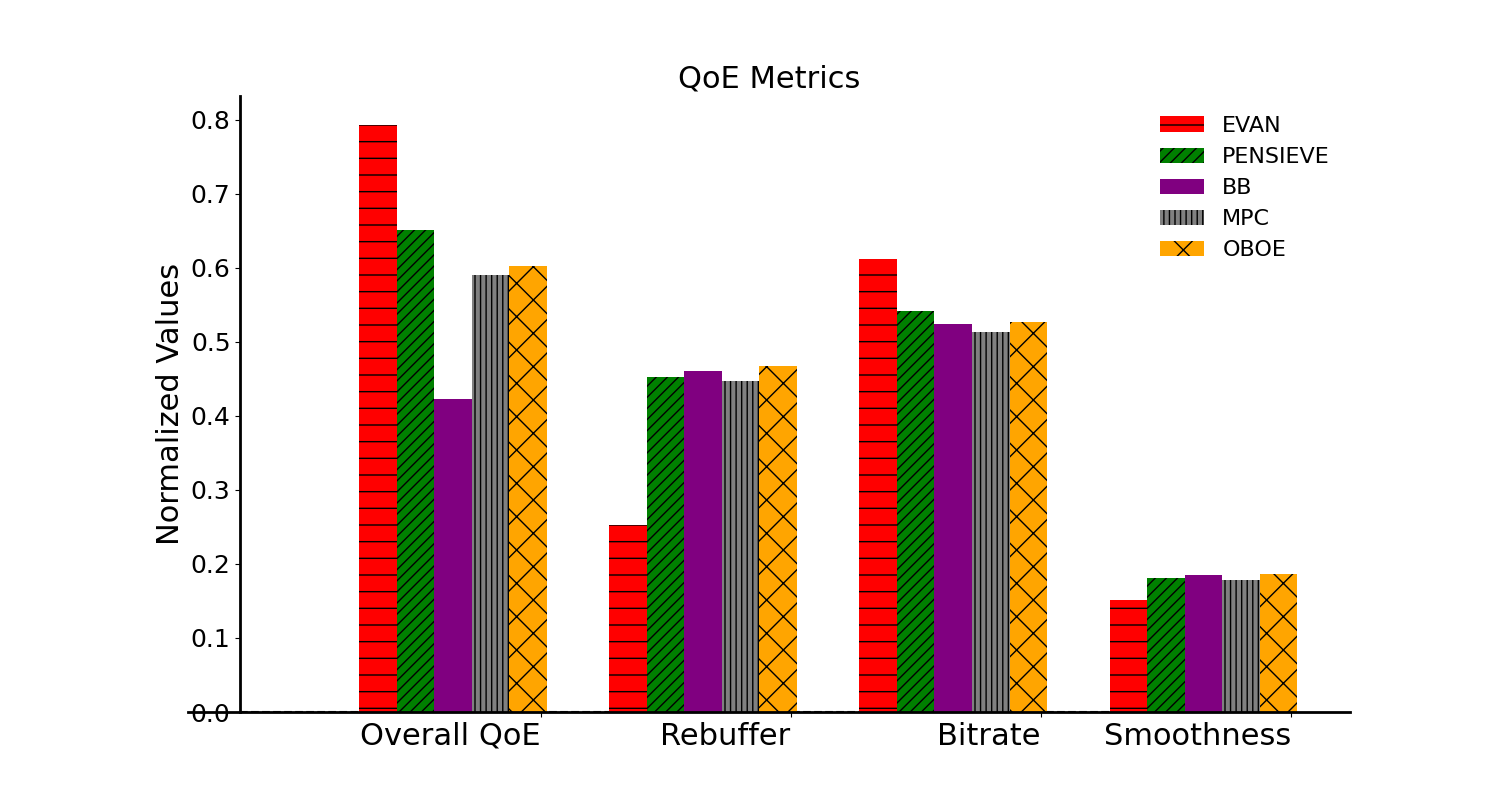}
     \caption*{(d)}
     \end{minipage}
     \begin{minipage}{0.5\linewidth}
     \centering
     \includegraphics[width=0.85\linewidth]{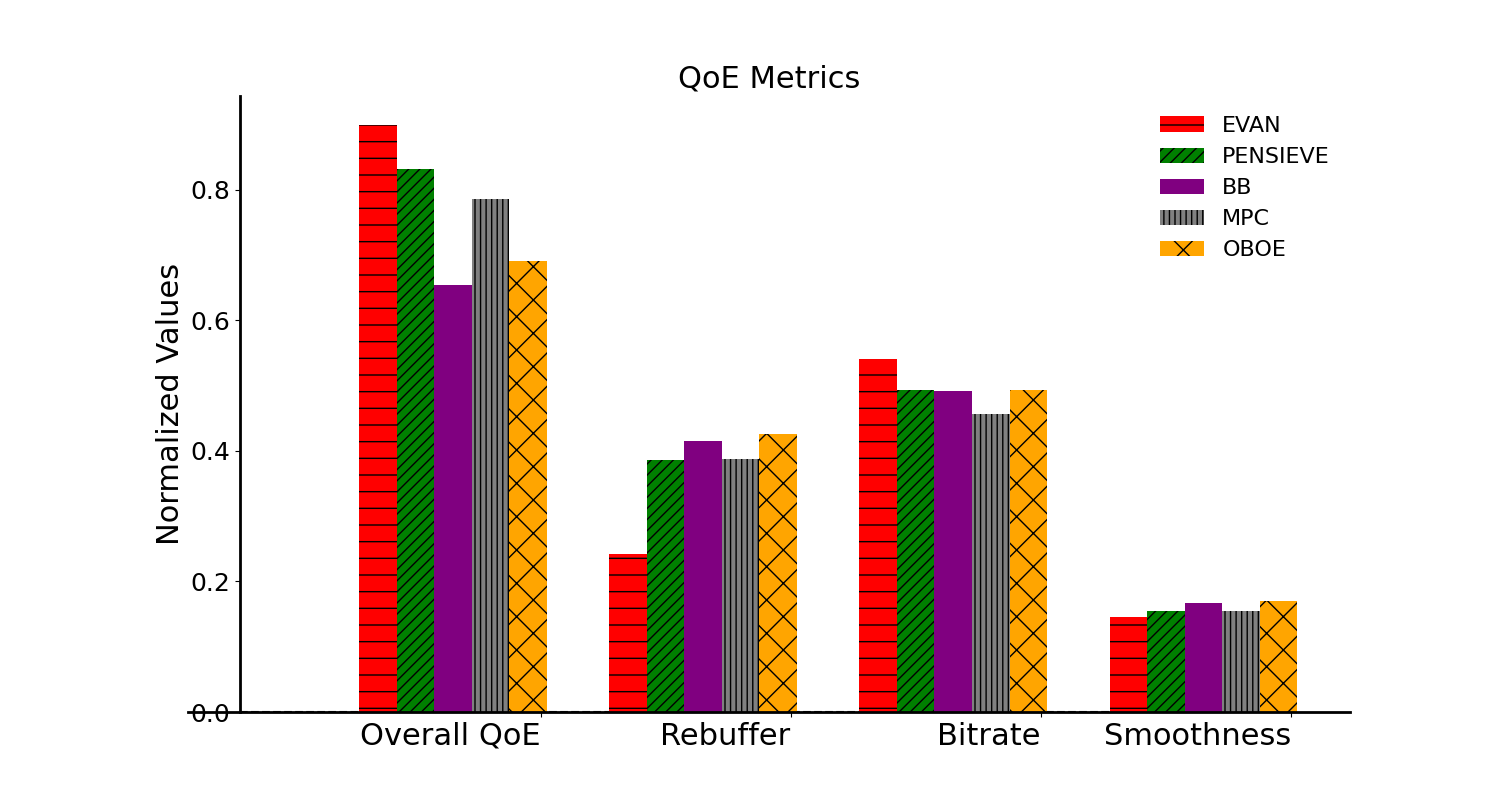}
     \caption*{(e)}
     \end{minipage}
     \caption{(a) PSNRs of H.264 and NeRV w.r.t BPP, prune ratio of NeRV model; (b) Total rewards for ABR evaluation on processed Norway \& FCC trace; (c) Total rewards for ABR evaluation on Oboe trace; (d) Total Rewards, rebuffering and video bitrate of the evaluated ABRs on processed Norway \& FCC trace; (e) Total Rewards, rebuffering and video bitrate of the evaluated ABRs on Oboe trace}
     \label{results}
\end{figure*}
\section{Experiments}
\begin{figure}[htbp]
\begin{minipage}{0.495\linewidth}
     \centering
     \includegraphics[width=1\linewidth]{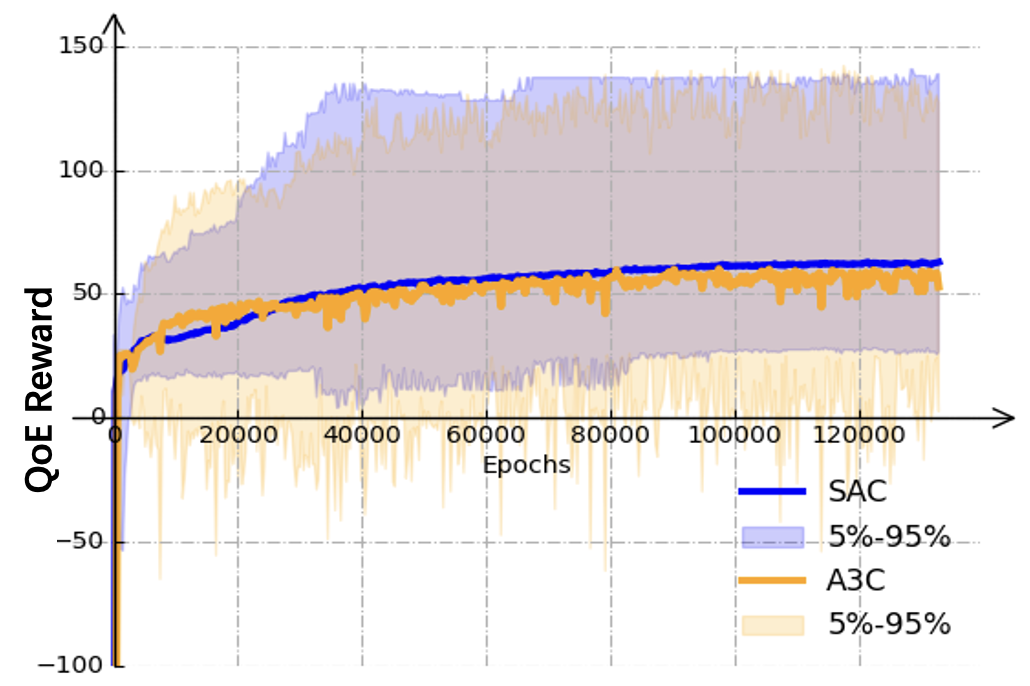}
     \caption*{(a)}
     \end{minipage}
     \begin{minipage}{0.495\linewidth}
     \centering
     \includegraphics[width=1\linewidth]{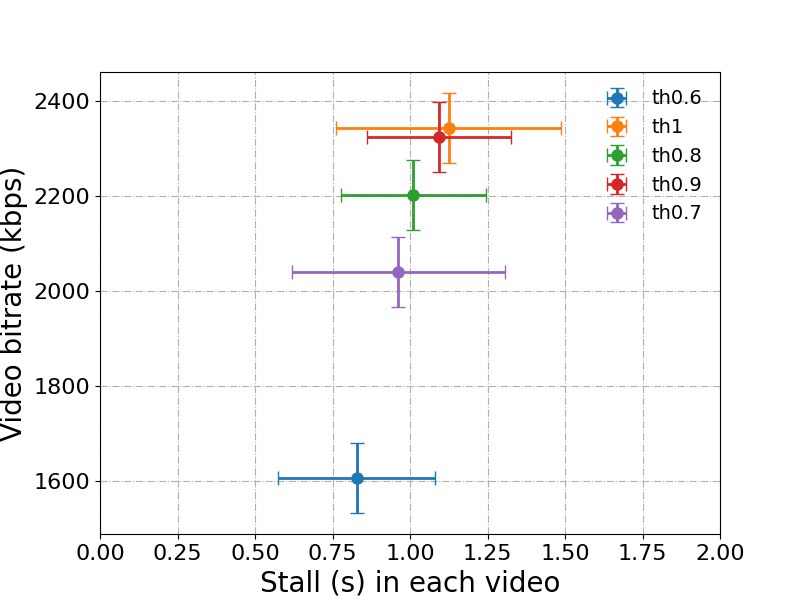}
     \caption*{(b)}
     \end{minipage}
     \caption{(a) Training rewards of SAC vs. A3C; (b) Video bitrate and Time stall of EVAN with different thresholds}
     \label{more}
\end{figure}

In this section, we experimentally assess EVAN’s performance. Our extensive evaluations answer the following questions: 1) How to construct different NeRV models for adaptive video streaming? 2) How does NeRV improve the total reward? 3) How does the playback threshold affect EVAN’s performance?
\paragraph*{\textbf{NeRV assessment}}
We employed the code from \cite{NeRV} to train and measure PSNRs with different BPPs and model sparsities, given video chunks in \cite{waterloo}. Each NeRV model was trained for 1200 epochs using Geforce RTX 3090. The results obtained, depicted in Fig. \ref{results} (a), reveal that NeRV can achieve a higher PSNR than the default codec H.264 for video streaming. We do not provide a performance comparison between NeRV and H.265, H.266, as H.264 remains the prevalent codec in current VoD systems. It is noteworthy that, \cite{NeRV} shows NeRV's performance remains superior to H.265 within a certain range of BPP. The PSNR curve for the pruned models suggests that incomplete models still show a remarkable video reconstruction quality even when more than 30\% of the weights are discarded (Fig. \ref{duration}). These two curves were then used to map the NeRV bitrates for QoE evaluation.
\paragraph*{\textbf{Dataset and Traces}}
We combined the pre-processed FCC and Norway traces with the Oboe trace \cite{oboe} to form our corpus for training and evaluation. To match the mean of the Oboe traces, we upscaled the FCC and Norway traces with Gaussian noise added to increase its mean from 2Mbps to 4Mbps. This is because the FCC and Norway traces were collected from older broadband and 3G networks, which are somewhat outdated for modern video streaming needs. We set the bitrate ladder as [750 Kbps, 1750 Kbps, 2350 Kbps, 3000 Kbps, 4300 Kbps, 7000 Kbps]. The setting is based on the representations provided in \cite{waterloo}. Additionally, we excluded the spatial resolution of 240p from the bitrate ladder since clients are generally unlikely to watch very low-quality representations. Unless otherwise specified, 80\% of our corpus are randomly taken to form the training set and the remaining 20\% for testing.
Each video from \cite{waterloo} were segmented into 4-second video chunks by \textit{FFmpeg}. To enable adaptive streaming via NeRV, we created NeRV models for each chunk representation of the same size.
Specifically, we vary the hyperparameters of NeRVs to create NeRVs that are capable of generating videos of different qualities. This is because higher-quality videos require more hidden features for reconstruction and vice versa. The hyperparameter configurations are listed in Table \ref{nerv model}.
\paragraph*{\textbf{EVAN v.s. other ABRs}}
We compared the performance of EVAN with the following benchmark methods:
\begin{itemize}
    \item MPC \cite{mpc}: A prevalent model-based ABR approach that utilizes a sliding window to maximize future chunks' reward
    \item Pensieve \cite{Pensieve}: A3C based reinforcement learning ABR method
    \item BB: A rule-based ABR based on buffer status
    \item OBOE \cite{oboe}: An ABR technique that adapts its strategy to different environments
\end{itemize}
\begin{table}
\centering
\caption{Simulation parameters}
\resizebox{0.38\textwidth}{!}{
\begin{tabular}{c|c|c|c}
\hline
\textbf{Parameter} & \textbf{Value} & \textbf{Parameter} & \textbf{Value}  \\
\hline
$\mathcal{H}_0$ (Initial entropy) & 1.78 & Rebuffer penalty & 4.3  \\
\hline
Training epochs & 120000 & Smooth penalty & 1  \\
\hline
Chunk Duration & 4s  & Chunks per test video clip & 40  \\
\hline
Actor learning rate & $10^{-4}$ & Maximum Buffer Size & 60 \\
\hline
Critic learning rate & $10^{-3}$  & State length & 8  \\
\hline
\end{tabular}}
\label{simulation}
\end{table}
We trained EVAN using A3C and SAC \cite{sac-git} for 12,0000 epochs with Tensorflow. The simulation parameters are summarized in Table \ref{simulation}. In the subsequent evaluation, we calculate the QoE for the entire video, which is the summation of all chunk's QoEs. As Fig. \ref{more} (a) shows, the SAC training process produces higher rewards with less variations. This is likely because SAC can adjust the entropy weight to achieve a better balance between exploitation and training, whereas A3C keeps the entropy fixed. Figs. \ref{results} (b) and \ref{results} (c) illustrate the average QoE achieved over the test traces. We found that EVAN outperformed all other ABRs in terms of the average QoE reward. The maximum average QoE increase was 22.4\% for Pensieve and 25\% for Oboe. 
\begin{table}
  \centering
    \caption{Intial re-buffering $T_{0}$ and buffer level $T$ (seconds)}
\resizebox{0.30\textwidth}{!}{
  \begin{tabular}{c|c c|c c}
    \hline
      & \multicolumn{2}{c|}{\textbf{Processed Norway FCC}} & \multicolumn{2}{c}{\textbf{Oboe}} \\
    \hline
    \textbf{ABRs} & $T_0$  $\downarrow$ & $T$ & $T_0$ $\downarrow$ & $T$ \\
    \hline
    \textbf{EVAN} & 0.51 & 8.52 & 0.62 & 7.47 \\
    \textbf{Oboe} & 0.93 & 10.75 & 0.87 & 10.27 \\
    \textbf{Pensieve} & 0.77 & 17.06 & 0.87 & 14.56 \\
    \textbf{MPC} & 1.01 & 12.38 & 0.86 & 10.13 \\
    \textbf{BB} & 1.03 & 8.80 & 0.97 & 9.12 \\
    \hline
  \end{tabular}}
  \label{rebuf}
\end{table}

Table \ref{rebuf} lists the initial re-buffering and buffer level of the ABRs in consideration. EVAN greatly reduces the average intial re-buffering by 33.7\% compared with Pensieve and maintains a lower buffer level for better download and playback tradeoff. Fig. \ref{results} (d) and \ref{results} (e) demonstrates the average QoE metrics. It is evident that EVAN has achieved a maximum of 50\% re-buffering reduction and has also an improved bitrate level. The reduction in re-buffering further confirms our initial motivation in designing EVAN. 
The results show that EVAN has significant performance improvement, mostly due to the higher bitrate level and re-buffering reduction that progressive playback offers. Since none of the compared schemes can 're-adjust' the bitrates once the decision has been made, EVAN, which is trained with SAC for higher bitrates and reduces re-buffering through progressive playback, can thus achieve enhanced performance with more flexible bitrate control.
\paragraph*{\textbf{Playback threshold}} 
As previously discussed, as the threshold was raised, the video bitrate increased, however, so did the re-buffering. The exponential relationship between PSNR and the model ratio indicated that excessively large thresholds for bitrate enhancement would not be sufficient to make up for the losses due to re-buffering. It is essential to pick a point that can significantly reduce re-buffering while not causing too much degradation in quality. EVAN was evaluated with different thresholds to roughly seek for the optimal point, as seen in Fig. \ref{more} (b). We noticed that when the threshold was shifted from 0.6 to 0.7, the bitrate increased by more than 400kbps, despite the fact that the re-buffering was increased by only 0.1 second. Conversely, when the threshold was increased from 0.7 to a higher value, the bitrate increase was not be enough to compensate for the re-buffering penalty, which led to a 4.3kbps bitrate reduction per re-buffering second in our QoE setting. Therefore, in the entire experiment, we set the threshold at 0.7.
\section*{Conclusion}
We proposed EVAN, an adaptive video streaming system that utilizes neural video representations (NeRV). This system is distinct from ABR approaches that use conventional codecs, as it is trained with a more effective strategy and progressive playback is used to prevent re-buffering. Thorough evaluations in various network conditions showed that EVAN can outperform existing ABR approaches by 8-21\%, and re-buffering was reduced by up to 50\%. Note that the training costs of NeRVs are still high. We shall look into the possibility of correlating weights with different contents complexity for weight sharing in order to reduce the training overhead in the future.

\bibliographystyle{IEEEtran}
\bibliography{IEEEabrv,Bibliography}

\end{document}